\documentclass[aip,reprint]{revtex4-1}

\usepackage{upgreek}
\usepackage{graphicx}
\usepackage{epstopdf}
\usepackage{color}

\draft % marks overfull lines with a black rule on the right

\begin{document}

\title{Resonance fluorescence from a telecom-wavelength quantum dot} 

% % % % % %Authors and Affiliations

\author{R. Al-Khuzheyri} \thanks{RA, ACD and JH contributed equally to this work.}%\email[]{ra328@hw.ac.uk} 
\affiliation{\mbox{Institute for Photonics and Quantum Sciences, Heriot-Watt University, Edinburgh EH14 4AS, UK}}

\author{A. C. Dada} %\email[]{a.c.dada@hw.ac.uk}
\thanks{RA, ACD and JH contributed equally to this work.}
\affiliation{\mbox{Institute for Photonics and Quantum Sciences, Heriot-Watt University, Edinburgh EH14 4AS, UK}}

\author{J. Huwer} %\email[]{jan.huwer@crl.toshiba.co.uk} 
\thanks{RA, ACD and JH contributed equally to this work.}
\affiliation{Toshiba Research Europe Limited, Cambridge Research Laboratory, 208 Cambridge Science Park, Milton Road, Cambridge CB4 0GZ, UK}

\author{T. S. Santana} %\email[]{t.s.santana@hw.ac.uk}
\affiliation{\mbox{Institute for Photonics and Quantum Sciences, Heriot-Watt University, Edinburgh EH14 4AS, UK}}

\author{ J. Skiba-Szymanska } 
\affiliation{Toshiba Research Europe Limited, Cambridge Research Laboratory, 208 Cambridge Science Park, Milton Road, Cambridge CB4 0GZ, UK}

\author{M. Felle} 
\affiliation{Toshiba Research Europe Limited, Cambridge Research Laboratory, 208 Cambridge Science Park, Milton Road, Cambridge CB4 0GZ, UK}
\affiliation{Centre for Advanced Photonics and Electronics, University of Cambridge, J. J. Thomson Avenue, Cambridge CB3 0FA, UK}

\author{M. B. Ward} %\email[]{martin.ward@crl.toshiba.co.uk}
\affiliation{Toshiba Research Europe Limited, Cambridge Research Laboratory, 208 Cambridge Science Park, Milton Road, Cambridge CB4 0GZ, UK}

\author{R. M. Stevenson}%\email[]{mark.stevenson@crl.toshiba.co.uk}
  \affiliation{Toshiba Research Europe Limited, Cambridge Research Laboratory, 208 Cambridge Science Park, Milton Road, Cambridge CB4 0GZ, UK} 

\author{I. Farrer}
\thanks{Present address: Department of Electronic \& Electrical Engineering, University of Sheffield, Sheffield S1 3JD, UK}
\affiliation{\mbox{Cavendish Laboratory, University of Cambridge, J. J. Thomson Avenue, Cambridge CB3 0HE, UK}}

\author{M. G. Tanner} %\email[]{m.tanner@hw.ac.uk} 
\affiliation{\mbox{Institute for Photonics and Quantum Sciences, Heriot-Watt University, Edinburgh EH14 4AS, UK}}

\author{R. H. Hadfield} %\email[]{Robert.Hadfield@glasgow.ac.uk} 
\affiliation{\mbox{School of Engineering, University of Glasgow, Rankine Building, Oakfield Avenue, Glasgow G12 8LT, United Kingdom}}

\author{D. A. Ritchie} %\email[]{jan.huwer@crl.toshiba.co.uk} 
\affiliation{\mbox{Cavendish Laboratory, University of Cambridge, J. J. Thomson Avenue, Cambridge CB3 0HE, UK}} 

\author{A.  J. Shields}%\email[]{jan.huwer@crl.toshiba.co.uk} 
 \affiliation{Toshiba Research Europe Limited, Cambridge Research Laboratory, 208 Cambridge Science Park, Milton Road, Cambridge CB4 0GZ, UK} 
 
\author{B. D. Gerardot} %\email[]{b.d.gerardot@hw.ac.uk} 
\affiliation{\mbox{Institute for Photonics and Quantum Sciences, Heriot-Watt University, Edinburgh EH14 4AS, UK}}

% % % % % %Authors and Affiliations

\date{\today}

\begin{abstract}
We report on resonance fluorescence from a single quantum dot emitting at  telecom wavelengths.   We perform high-resolution spectroscopy  and  observe the Mollow triplet in the Rabi regime---a hallmark of resonance fluorescence. The measured resonance-fluorescence spectra  allow us to rule out pure dephasing as a significant decoherence mechanism in these quantum dots. Combined with numerical simulations, the experimental results provide robust characterisation of charge noise in the environment of the quantum dot.  Resonant control of the quantum dot  opens up new possibilities for  on-demand  generation  of  indistinguishable single photons  at telecom wavelengths as well as  quantum optics experiments  and direct  manipulation of solid-state qubits in telecom-wavelength quantum dots.
\end{abstract}

%\pacs{68.65.Hb}   

\maketitle

%\section{Introduction}

\begin{figure}
\centering
\includegraphics[width=0.9\linewidth]{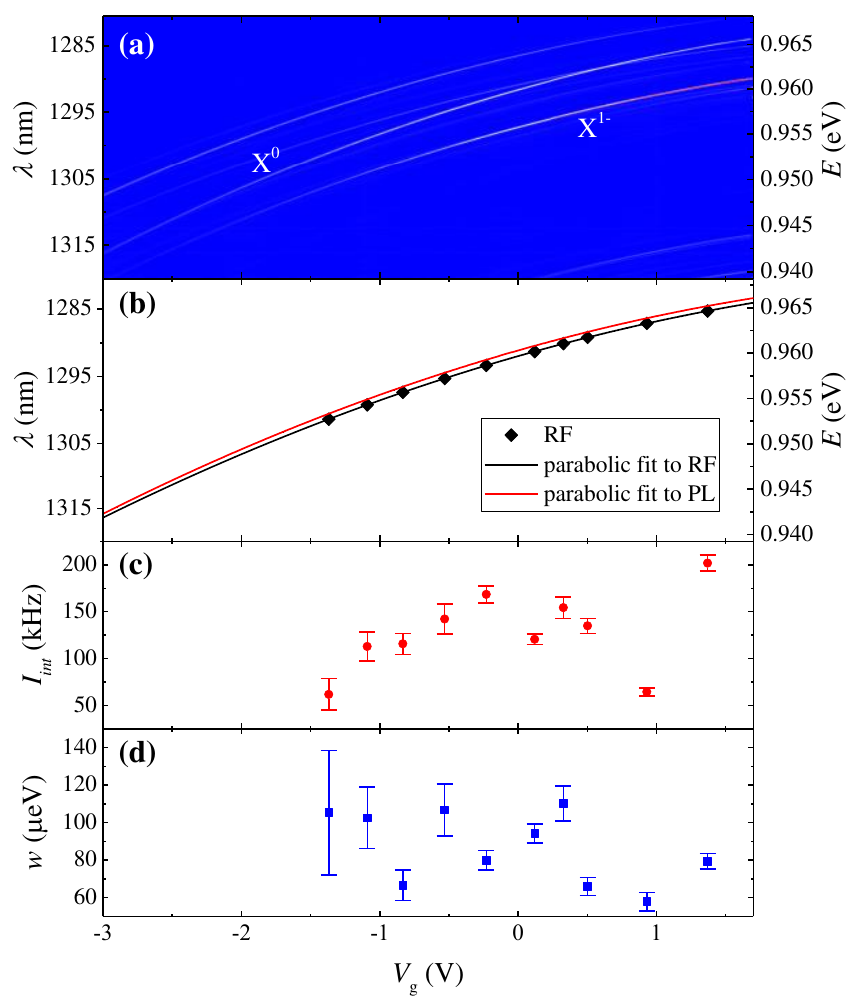}
\caption{(a) {\bf Microphotoluminescence-voltage map. }  Spectra collected as a function of the applied diode bias from a single QD within the intrinsic region of a \textit{p-i-n} diode under excitation at a wavelength of 1060~nm. The two lines of interest correspond to emission from the neutral and charged exciton from a single QD.  
{\bf Exciton plateau mapping in resonance fluorescence}. (b) Resonance energies  showing Stark shifts with permanent dipole moment $p/e=0.420(4)$~nm and  polarizability $\beta=-0.2140(1)$~$\upmu$eV/(kV/cm)$^2$ for RF. PL is plotted for comparison. (c) integrated RF counts ($I_{int}$) and (d) FWHM of Lorentzian fits to voltage-detuning RF spectra taken at different excitation wavelengths at a power of 77~nW ($\Omega=409$~MHz).}
\label{fig:Fig1}
\end{figure}

\begin{figure*}
\centering
\includegraphics[width=0.925\linewidth]{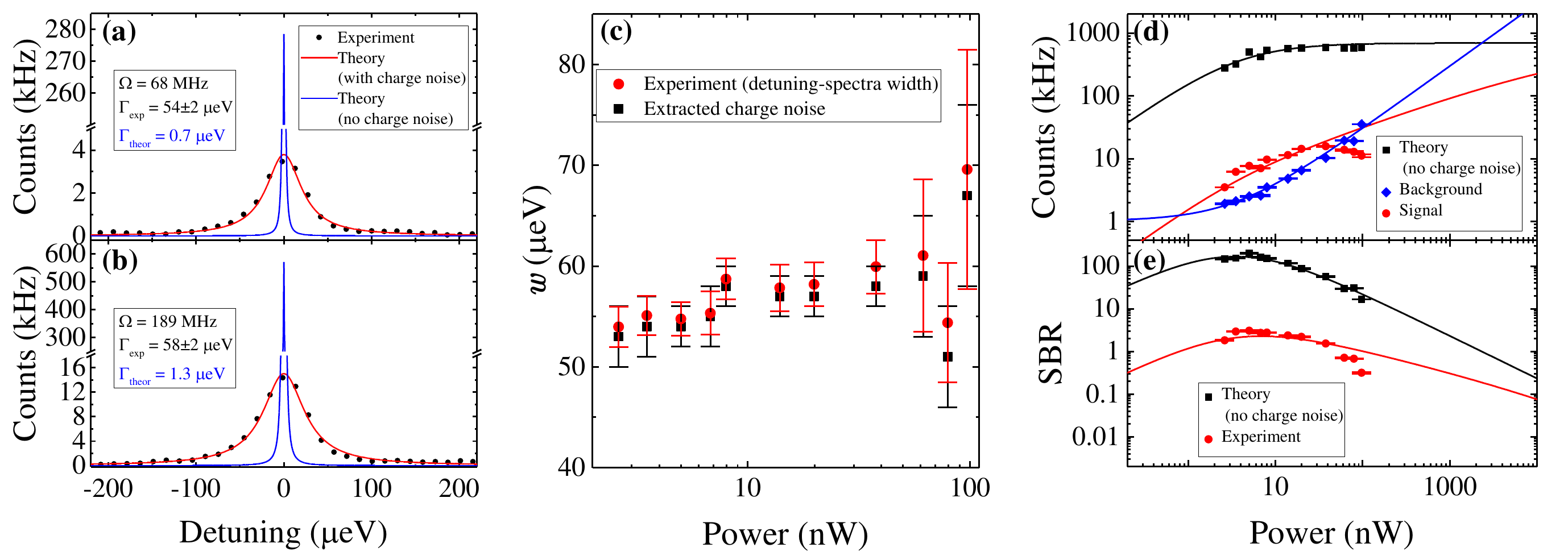}
\caption{(a),(b) {\bf Resonance fluorescence detuning spectra} collected from the QD $X^-$ state with the laser background subtracted. Detuning spectra obtained analytically using the master equation method with the charge noise parameter free was fitted to the data (red) and the case with no charge noise (blue) plotted for comparison. The scans were performed for diode bias voltages 1.3-1.45\hspace{1pt}V in 5-mV steps. (c) {\bf Linewidths of detuning spectra and extracted charge noise. }  Red dots: overall linewidth observed in detuning spectra; black squares: corresponding charge noise characteristic width. (d), (e)  {\bf Power dependence of resonance fluorescence counts}. The amplitude of the Lorentzian  fits to the background-subtracted RF voltage detuning spectra  at $ \lambda=1285.28200$\hspace{1pt}nm are plotted as a function of excitation power. Represented are background-subtracted signal (red dots), background counts  measured off resonance (blue dots), charge-noise-corrected experimental data (black squares) and theoretical saturation curves for the corresponding cases (solid lines). (e) shows SBR power dependence.
}
\label{fig:Fig2}
\end{figure*}

\begin{figure*}
\centering
\includegraphics[width=0.96\linewidth]{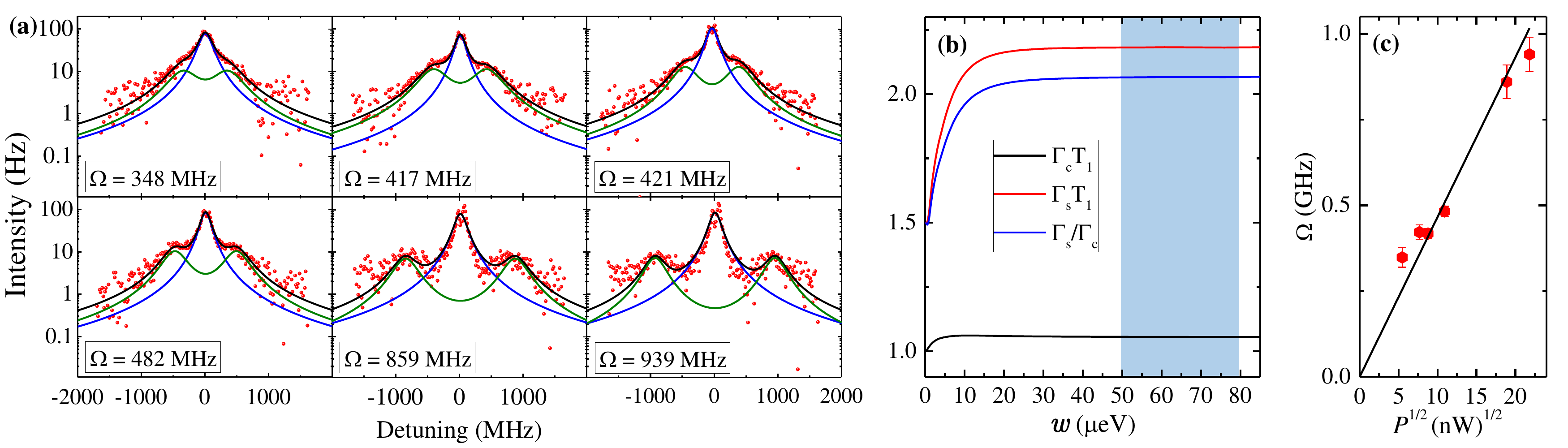}
\caption{{\bf (a)  Observation of Mollow triplet in high-resolution telecom resonance-fluorescence spectra.} The inelastically scattered component of the spectra is shown for six different excitation powers corresponding to the indicated Rabi frequencies $\Omega$. The plots show  the QD $X^0$ RF  filtered through the FPI as a function of FPI detuning (red dots). Lorentzian fit components: the side peaks (green curve), the central peak (blue curve), and the total (black curve). {\bf (b)} Theoretical simulation of FPI-spectra line-widths (FWHM) of the central ($\Gamma_c)$  and either of the side peaks ($\Gamma_s$) as a function of $\Omega$ obtained from numerical simulation of the master equation.  We show the respective ratios for linewidths and comparison with the transform limited case $\Gamma {\rm T}_1=1$. The shaded region indicates the regime of our measurements. {\bf (c)} The Rabi splitting/frequency as a function of the square root of excitation power showing the expected linear dependence. We are unable to extract precise values for charge noise contributions from the the FPI data in (a)  because the measurements lie in the shaded region in (b).}
\label{fig:Fig3_fpi}
\end{figure*}

Resonance fluorescence (RF) is a result of coherent interaction between an electromagnetic field and a two-level atomic system~\cite{loudon2000quantum,scully1997quantum}. The ability to access the emission resulting from resonant excitation of a quantum dot (QD)~\cite{Muller:2007ek}  has been shown to furnish unique possibilities to investigate intriguing quantum optical effects in solid-state systems, such as non-classical light generation manifesting near-ideal antibunching\cite{Ding:2016cy,loredo2016scalable,Somaschi:2016ej}, 
 entanglement~\cite{Muller:2014ir}, squeezing~\cite{Schulte:2015kr},  as well as quantum interference and Rabi-oscillations~\cite{Schaibley:2013bt,dada2016indistinguishable}. 
  These can, in turn, be harnessed for quantum information science.   
 Quantum communication networks having nodes consisting of stationary matter qubits interconnected by flying photonic qubits~\cite{Delteil:2016fx} will require telecom photons to enable long-distance applications. This makes a coherent interface  of quantum dot spins with telecom wavelength photons particularly desirable.   
   Also, due to considerably reduced decoherence effects under resonant excitation,  the best single-photon sources in terms of demonstrated indistinguishability, purity, on-demand operation and brightness, are based on QD RF~\cite{Ding:2016cy,loredo2016scalable,Somaschi:2016ej,dada2016indistinguishable}.
    While QD single-photon emission occurs under incoherent pumping provided by nonresonant excitation, the latter also generates extra carriers in the host material. This leads to inhomogeneous broadening of the emission from spectral wandering due to charge fluctuations \cite{Houel:2012bp, Kuhlmann:2013era,Matthiesen:2014gl}, as well as time jitter between photon absorption and emission due to an uncontrolled step of non-radiative relaxation to the exciton state before recombination~\cite{Santori:2004fr}, making RF preferable for applications exploiting single photons. 

Despite the promise that telecom-QD RF offers, the body of previous work demonstrating and applying QD RF has been limited to QDs emitting at $<1\mu\hspace{1pt}$m, where losses in silica fiber are high such that long-distance quantum communication protocols~\cite{Kimble:2008if} become unfeasible.   In particular, QDs emitting at $\lambda\sim950$\hspace{1pt}nm have been demonstrated extensively as bright sources of coherently-generated indistinguishable single photons~\cite{loredo2016scalable,Huber:2015uj,dada2016indistinguishable,Ding:2016cy,Somaschi:2016ej,Matthiesen, PhysRevLett.103.167402, PhysRevB.90.041303,bennett2015cavity}  
 and as a bright source of entangled photon-pairs~\cite{Muller:2014ir} via resonant excitation. 
Translating this  progress to QDs emitting in the telecommunication O-band ($\lambda\sim1310$\hspace{1pt}nm) or C-band ($\lambda\sim1550$\hspace{1pt}nm) has proved challenging.  
Also, while QD spin-photon entanglement has been demonstrated~\cite{Schaibley:2013jb,gao2013quantum}, including the extra complexity of downconversion to telecom wavelengths~\cite{DeGreve:2012gec}, actual observation of fluorescence due to resonant interaction between a telecom-wavelength photon and a quantum dot is  yet to be reported.  Even so,  studies of telecom QDs under nonresonant excitation have characterised their confinement and spin properties~\cite{Sapienza:2016ij} and demonstrated their potential as sources of single indistinguishable \cite{Kim,Felle:2015ef} and entangled photons~\cite{Ward}.  Here, we demonstrate RF from a single QD emitting in the telecom O-band. In spite of considerable charge noise in the environment of the QD, we observe nearly transform-limited linewidths for the central incoherent peak of the Mollow triplet. This signifies negligible pure dephasing in the QD.

The device was grown on a GaAs substrate by molecular beam epitaxy. The InAs-QD layer lies in a quantum well located within the intrinsic region of a \mbox{\emph{p-i-n}} diode. The device contains a weak planar cavity consisting of  AlGaAs/GaAs distributed Bragg reflectors for improved collection efficiency. The QDs are located in  3-$\upmu$m-diameter apertures in a 100-nm-thick evaporated Al layer covering the surface of the device to facilitate navigation to individual QDs. A  similar QD has been described in a previous study based on nonresonant excitation~\cite{Ward}. Measurements were performed at \mbox{4\hspace{1pt}K} using a high-numerical-aperture confocal-microscope setup 
for single-QD spectroscopy. The scattered laser light was suppressed by using  
 orthogonal linear polarizers in the excitation and collection arms of a confocal microscope 
 with  typical extinction  ratios $>$$10^6$.  
A  tuneable continuous-wave (CW) laser diode was used for resonant excitation. For photon counting, we used both a NbTiN superconducting-nanowire single-photon detectors (SNSPD)~\cite{Tanner:2010kn}  and a cooled InGaAs single-photon avalanche diode (SPAD) having a peak efficiency of $\sim$$20\% $ at $\sim$$1310$\hspace{1pt}nm for benchmarking. 

First, we acquire a micro-photoluminescence ($\upmu$-PL) map with a 70-$\mu$eV-resolution  grating spectrometer by varying the \emph{p-i-n} diode bias voltage under non-resonant excitation  with a diode laser at $\lambda\sim1060$\hspace{1pt}nm.  In Fig.~\ref{fig:Fig1}(a), the map clearly shows the QD neutral ($X^0$) and charged ($X^-$) exciton lines.  This characterisation allows the tuning of the resonant-excitation laser to the desired QD  transition for RF measurements. Further, we obtain the lifetime for $X^{0}$  from time-resolved $\mu$-PL measurements to be $ T_1=1.394(6)$\hspace{1pt}ns. As measured using a Mach-Zehnder interferometer, the total coherence time under non-resonant excitation is $ T_{\rm 2,nres}=89(8)$\hspace{1pt}ps.  We base our RF investigations on $X^0$.

 By obtaining detuning spectra at  various excitation wavelengths, we map the RF signal over the $X^0$ plateau at a fixed power above saturation. The integration time for each point in the RF measurements is 5\hspace{1pt}s. Fig.~\ref{fig:Fig1}(b) shows the peak energies as a function of \emph{p-i-n} voltage while  Fig.~\ref{fig:Fig1}(c) shows the integrated counts obtained from the fits to the RF detuning-spectra data acquired for a series of excitation wavelengths. 
The linewidth over the extent of the $X^0$ plateau [shown in  Fig.~\ref{fig:Fig1}(d)] tends to decrease at more positive voltages/longer wavelengths. 
 The $X^0$ plateau mapping via RF manifests a clear Stark shift. 
 The dependence of peak energies on the electric field $F$ (via the \emph{p-i-n} diode bias voltage $V_g$)  is  $E_{PL}=E_{0}-pF+\beta F^{2} $.  The field, which is a function of $V_{0}$ and the thickness of the intrinsic region ($d$),  is $F=-(V_{g}-V_{0})/{d}$. 
In this case, $V_{0}=2.2$\hspace{1pt}V and $d=203\hspace{1pt}$nm \cite{Ward}. The permanent dipole moment ($p$) and the polarizability ($\beta$) are extracted from the fit as $p/e=0.420(4)$~nm and $\beta=-0.2140(1)$~$\upmu$eV/(kV/cm)$^2$. For comparison, the nonresonant-excitation  (1060-nm) case gives $p/e=0.385(1)$~nm and $\beta=-0.2290(4)$~$\upmu$eV/(kV/cm)$^2$. We understand these small differences to be due to additional charging of the host semiconductor matrix induced by nonresonant optical excitation.
 
The data points in Fig.~\ref{fig:Fig2}(a) and (b) show background-subtracted RF counts as we tune the $X^{0}$ transition through resonance with the laser using the \emph{p-i-n} diode bias voltage.  We show two examples with the resonant excitation laser and voltage across the diode set to \mbox{$\lambda=1285.28200$~nm} and $V_{\rm g}=1.371$\hspace{1pt}V respectively [due to reduced linewidths in detuning spectra, see Fig.~\ref{fig:Fig1}(a),(d)]  at powers respectively below and above saturation.   Fig.~\ref{fig:Fig2}(c) shows the linewidths obtained at different excitation powers from the detuning spectra.  We demonstrate that the RF counts manifest saturation behaviour in Fig.~\ref{fig:Fig2}(d) and (e) which respectively show the (background-subtracted) RF counts plotted with the background counts and the corresponding signal-to-background ratios (SBR). 
 We observe SBRs of $\lesssim3$ [Fig.~\ref{fig:Fig2}(e)], and suspect that background counts are  mainly due to  scattering of the excitation laser light off the structured sample surface.  
We note briefly that the characteristic double peak for $X^0$ corresponding to its fine-structure splitting (FSS $= 109(4) \upmu$eV at $V_{\rm g}= 1.371 V$ in non-resonant PL)  is not observable in these RF detuning spectra, possibly due to the diminished visibility of a smaller-intensity line  by the low SBR.

We perform high-resolution  spectroscopy of the RF from the QD using a Fabry-P\'erot scanning interferometer (FPI) with a 5-GHz free-spectral range and 33-MHz resolution. The integration time for each FPI measurement is $\sim$$20$\hspace{1pt}mins.
 The inelastically scattered component of the acquired spectra, shown in Fig.~\ref{fig:Fig3_fpi}(a), clearly reveals the Mollow triplet~\cite{Mollow:1969dd}, which is a quintessential feature of RF from a two-level system. We subtracted the narrow-linewidth elastic peak~\cite{Matthiesen:2012je,Konthasinghe:2012gp,Nguyen:2011dh} which was contaminated by background laser signal. The overall fit (black line) to the data (red dots) is the sum of three Lorentzian functions that fit the two inelastic side peaks (green curve) and the inelastic central peak (blue curve).  
  
Finally, we demonstrate the use of RF to probe the effect of charge noise~\cite{Houel:2012bp,Kuhlmann:2013era,Matthiesen:2014gl} in the telecom-wavelength QD sample.  Charge noise  at timescales much longer than $T_1$ and $T_2$ manifests in  the voltage-detuning spectra as a characteristic broadening, while in the high-resolution FPI spectra, it primarily broadens the sidebands of the Mollow triplet. We model this effect of charge noise by calculating the RF spectra for a 2-level system using the master-equation method~\cite{loudon2000quantum,scully1997quantum}, including the effects of dephasing due to spontaneous emission, pure dephasing, and slow charge noise. This enables us to obtain an analytical result for the detuning spectra with which we fit the data [e.g.,  in Figs~\ref{fig:Fig2}(a),(b)]. In modelling charge noise, we assume $T_1<<T_c<<T_{\rm exp}$~\cite{Konthasinghe:2012gp,Kuhlmann:2013era,Matthiesen:2014gl}, where $T_{\rm exp}$ is the experiment acquisition time, and $T_c\sim1$\hspace{1pt}ms is the timescale of spectral fluctuations due to charge noise. These measured spectra reveal linewidths much greater than transform-limited linewidths based our measured $T_1$. We quantify the charge noise by a characteristic width parameter $w_c$ which is aproximately equal to the  width of the detuning spectra $w$ [see Fig.~\ref{fig:Fig2}(c)], since $w_c>>1/T_1$.  By simulating the high-resolution RF spectra we obtain numerical results corresponding to various amounts of charge noise $w$.  We find that the width of the central peak in our Mollow-triplet data [$\Gamma_{c,{\rm min}}=0.124(5)\hspace{1pt}$GHz] is consistent with the case of negligible pure dephasing (i.e. $T_{\rm 2, res}\approx 2 T_1 \approx 2.6\hspace{1pt}{\rm ns}>> T_{\rm 2, nres}\approx 90\hspace{1pt}$ps ), which was then assumed for the fitting of the detuning spectra [Fig.~\ref{fig:Fig2}(a),(b)]. This further demonstrates that the time scale of spectral fluctuations due to charge noise $T_c>>T_1$.

In summary, we present an experimental demonstration of RF from a single QD emitting at telecom wavelengths ($\lambda\approx1300$~nm). 
We observe Mollow triplet emission--a key signature of RF, with Rabi splitting showing the expected square-root dependence on excitation power, and demonstrate a contrast in dephasing times between resonant and non-resonant excitation.  Crucially, the near transform-limited linewidths observed in the RF spectra confirm negligible pure dephasing in these telecom wavelength QDs. We also characterise the charge noise in our sample using RF. 
In future work, we expect that charge noise can be minimised through improved sample design and fabrication.   
The results pave the way for directly interfacing stationary matter qubits with telecom wavelength photons, highly-coherent single-photon emission,  on-demand generation of indistinguishable photons and polarization-entangled photon pairs via resonance fluorescence at telecom wavelengths.  

{Acknowledgements}: 
This work is funded by the EPSRC (grant numbers EP/I023186/1, EP/K015338/1, EP/M013472/1, EP/G03673X/1, EP/K0153338/1, EP/I036273/1 and EP/J007544/1) and an ERC Starting Grant (number 307392). BDG and RHH acknowledge  Royal Society University Research Fellowships. RHH thanks Prof. V. Zwiller and Dr. S. Dorenbos for providing the SNSPD detector chip.

%========Bibliography========
%========Bibliography========

%\bibliographystyle{aipnum4-1}
%\bibliography{TelecomRF_refs}

%========Bibliography========
%========Bibliography========


%========Bibliography========

\end{document}